\documentclass[preprint,prd,showpacs,nofootinbib]{revtex4}

\usepackage{doi}
\usepackage{hyperref}
\hypersetup{
  colorlinks=true,        
  linkcolor=blue,         
  citecolor=cyan,         
}

\usepackage{bm}
\usepackage{latexsym}
\usepackage{dcolumn}
\usepackage{amsmath,amsfonts,amssymb}
\usepackage{amsthm}
\usepackage{color}
\usepackage{xcolor}
\usepackage{graphicx,float}
\usepackage{comment}
\usepackage{hyperref}

\usepackage{pdfpages}
\usepackage{epstopdf}
\usepackage[utf8]{inputenc}

\newtheorem{remark}{Remark}

\def\be {\begin{equation}}
\def\ee {\end{equation}}
\def\bea {\begin{eqnarray}}
\def\eea {\end{eqnarray}}
\def\bc {\begin{center}}
\def\ec {\end{center}}
\def\bfg {\begin{figure}}
\def\efg {\end{figure}}
\def\bi {\begin{itemize}}
\def\ei {\end{itemize}}

\DeclareMathOperator{\SU}{SU}
\DeclareMathOperator{\U}{U}
\DeclareMathOperator{\GL}{GL}
\DeclareMathOperator{\SL}{SL}
\DeclareMathOperator{\M}{M}

\def\beq{\begin{equation}}
\def\eeq{\end{equation}}
\def\br{\begin{eqnarray}}
\def\er{\end{eqnarray}}
\newcommand{\eel}[1] {\label{#1}\end{equation}}

\newcommand{\bdm}{\begin{displaymath}}
\newcommand{\edm}{\end{displaymath}}

\begin{document}


\title{Unitary symmetries in wormhole geometry and its thermodynamics}

\author{Ahmed~Farag~{\bf A}li$^{\triangle \nabla}$}
\email{aali29@essex.edu ;  ahmed.ali@fsc.bu.edu.eg}
\author{Emmanuel {\bf M}oulay$^\otimes$}
\email{emmanuel.moulay@univ-poitiers.fr}
\author{Kimet {\bf J}usufi$^\odot$}
\email{kimet.jusufi@unite.edu.mk}
\author{ Hassan {\bf A}lshal$^{\bigcirc\boxdot}$}
\email{halshal@scu.edu}
\affiliation{\footnotesize{$^\triangle$Essex County College, 303 University Ave, Newark, NJ 07102, United States.}}
\affiliation{\footnotesize{$^\nabla$Department of Physics, Faculty of Science, Benha University, Benha, 13518, Egypt.}}

\affiliation{\footnotesize{$^\otimes$XLIM (UMR CNRS 7252), Universit\'{e} de Poitiers, 11 bd Marie et Pierre Curie, 86073 Poitiers Cedex 9, France}}

\affiliation{\footnotesize{$^\odot$Physics Department, State University of Tetovo,
Ilinden Street nn, 1200, Tetovo, North Macedonia}}

\affiliation{\footnotesize{$^\bigcirc$Department of Physics, Santa Clara University,  500 El Camino Real, Santa Clara, CA 95053, United States}}

\affiliation{\footnotesize{$^\boxdot$Department of Physics, Faculty of Science, Cairo University, Giza 12613, Egypt}}

\begin{abstract}
\noindent
 From a geometric point of view, we show that unitary symmetries $\U(1)$ and $\SU(2)$ stem fundamentally from Schwarzschild and Reissner-Nordström wormhole geometry through spacetime complexification. Then, we develop quantum tunneling which makes these wormholes traversable for particles. Finally, this leads to wormhole thermodynamics.

\begin{center}
``As Above So Below'' THOTH\\
\end{center} 

\end{abstract}

\maketitle

   \tableofcontents
\section{Introduction}
\noindent 
Einsiten-Rosen wormhole was introduced to understand the geometric meaning of mass and charge of the elementary particles in Ref.
 \cite{einstein1935particle} and then was developed by many authors \cite{abbott1989wormholes,barcelo1999traversable,hawking1987quantum,jusufi2018conical,kruskal1960maximal,lobo2017wormholes,morris1988wormholes,morris1988wormholes2,visser1990quantum}.  The geometric description of physical concepts was a cornerstone of several approaches to quantum gravity. These approaches include noncommutative geometry \cite{van2015noncommutative}, string theory \cite{zwiebach2009first}, loop quantum gravity \cite{gambini2011first} and twistor theory \cite{huggett1994introduction}. In this article, we focus our attention on a fundamental question: \emph{is there a conceptual connection between unitary symmetries and wormhole geometry}? We argue that it is possible to find unitary symmetries, such as $\U(1)$ and $\SU(2)$, from Schwarzschild and Reissner-Nordström wormhole geometry through spacetime complexification if a new Euclidean metric on a complex Hermitian manifold is provided. This motivates us to compute quantum tunneling, which indicates that these wormholes could be traversable for particles. Finally, this allows us to introduce wormhole thermodynamics that is consistent with black hole thermodynamics \cite{bekenstein1972black,hawking1975particle}.\\
\noindent
The article is organized as follows. We start with the Schwarzschild wormhole geometry in Section~\ref{Sec:Schwarzschild}, and we connect its complex geodesics with $\U(1)$ and $\SU(2)$ symmetries by using spacetime complexification. We also provide a new Euclidean metric on a Hermitian complex manifold. In Section~\ref{Sec:RN}, the massless exotic Reissner-Nordström wormhole geometry is also connected with the same unitary symmetries, and a discussion about the classical Reissner-Nordström wormhole geometry and the $\SU(3)$ symmetry is addressed. Quantum tunneling for particles is studied in Section~\ref{Sec:tunneling} and lead to wormhole thermodynamics. Finally, concluding remarks are given in Section~\ref{Sec:conclusion}.

\section{Schwarzschild wormhole geometry}\label{Sec:Schwarzschild}
\begin{figure}[H]
    \centering
    \includegraphics[width = 0.5\textwidth]{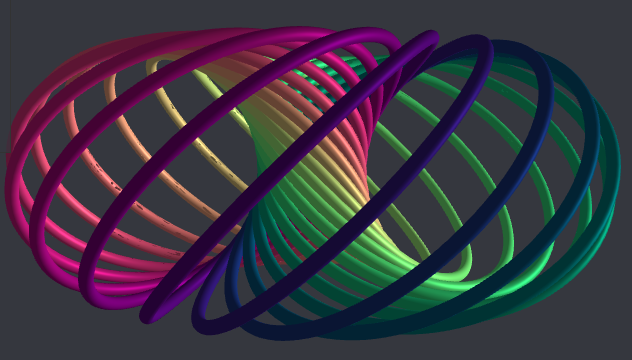}
    \caption{Hopf fibration of $S^3$ (see \url{https://philogb.github.io/page/hopf/})}
    \label{fig:Hopf}
\end{figure}

\noindent
It is historically known that Einstein and Rosen (ER) introduced their ER bridge, or wormhole idea, to resolve the particle problem in General Relativity (GR) \cite{einstein1935particle}. The ER bridge contrives a geometric meaning of particle properties, such as mass and charge, in the spacetime, where mass and charge are nothing but bridges in the spacetime. The ER bridge idea can be summarized as follows. The Schwarzschild metric is given by
\be
ds^2= -\left({1-\frac{2M}{r}}\right)^{-1} dr^2- r^2\left(d\theta^2+\sin^2\theta d\phi^2\right) +\left(1-\frac{2M}{r}\right)dt^2~.
\ee
where $M>0$. It has both the physical singularity existing at $r=0$ that cannot be removed, and the coordinate singularity at $r=2M$ that can be removed by choosing another coordinate system. Einstein and Rosen suggested a coordinate system which resolves the coordinate singularity at $r=2M$ by choosing the following transformation
\be\label{u2=r-2m}
u^2=r-2M~,
\ee
leading to $4u^2du^2=dr^2$. In the new coordinate system, one obtains for $ds^2$ the expression
\be\label{metric_S1}
ds^2=-4 (u^2+2M)du^2-(u^2+2M)^2(d\theta^2+\sin^2\theta d\phi^2)+\frac{u^2}{u^2+2M}dt^2~.
\ee
\noindent
One may notice in this coordinate system that $u$ will be real value for $r> 2M$ and will be imaginary for $r< 2M$. As $u$ varies from $-\infty $ to $\infty $, one finds $r$ varies from $+\infty$ to $2M$ and then from $2M$ to $+\infty$. In that sense, the $4-$dimensional spacetime can be described by two congruent sheets that are connected by a hyperplane at $r=2M$, and that hyperplane is the so-called ``bridge''. Thus, Einstein and Rosen interpreted mass as a bridge in the spacetime.
\\
This draws our attention to look closely at the case when $r<2M$, and consequently the variable $``u"$ would have imaginary values in this region. The geodesics in the $u-$coordinate system will experience two different kinds in two different regions. In the region $r>2M$, it would follow real trajectory, and it follows imaginary trajectory in the region $r<2M$. But as we cannot ``stitch'' a real space and a complex space together, we prefer to \textit{complexify} the whole spacetime.
It might be enticing to impose real spatial indices to formulate complex geodesics as the spatial coordinate $u$ is what motivates us to consider { \emph{spacetime complexification}}, but the more wise choice is to cook one complex dimension from spatiotemporal dimensions and the other complex dimension from the leftover spatial dimensions. Also, we classify the geodesics based on real and imaginary parts in the two sheets of the wormhole. This is crucial to develop a more consistent theory of gravity for the following reasons:
\begin{itemize}
\item The manifold in GR is chosen to be pseudoRiemannian manifold \cite{gromov1980classification}, which is connected and guarantees the general covariance and continuous coordinate transformations on the manifold. But a basic question emerges: To what does the region $r<2M$ develop under diffeomorphisms? The answer should include that there must be a geometric structure, by covarience principle, that corresponds to the region in wormhole geometry.
\item The physical singularity at $r=0$ is irremovable by coordinate transformation in GR \cite{hawking1970singularities}, which implies the importance of studying the region connected with $r=0$, even in the coordinates that give wormholes, as it is likely to have a correspondence in wormhole geometry.
\item In wormhole geometry, the $u$ values become imaginary for $r<2M$. Imaginary value in physics plays crucial rule in building unitary symmetries. We are interested to understand the effect of this imaginary region in wormhole geometry knowing that the role of complex numbers in QM is recognized as to be a central one \cite{hickey2018quantifying}.
\end{itemize}
\noindent
In order to complexify a spacetime $\mathcal{N}$, or to think of $\mathcal{N}\rightarrow\mathbb{R}^4$ as $\mathcal{M}\rightarrow\mathbb{C}^2$, we introduce complex manifold $\mathcal{M}$ of two complex dimensions $\zeta$ and $\eta$. We consider a point $p \in \mathcal{M}$ so that $p = (\zeta, \eta)$ defines the complex coordinates in some local chart
\begin{subequations}\label{sheets C-transformation}
\begin{align}
\zeta= \zeta_1+ i \zeta_2~,\\
\eta= \eta_1+ i \eta_2~,
\end{align}
\end{subequations}
where the complex coordinates induce the parameter space of the real parameters $(\zeta_1,\eta_1,\zeta_2,\eta_2)$ on $\mathcal{M}$.
In that sense, the full geodesic in wormhole geometry would read 
\begin{subequations}
\begin{align}
\lambda_1(\zeta_1,\eta_1,\zeta_2,\eta_2)= \lambda_1(\zeta_1, \eta_1)\in\mathbb{R}~,\\
\lambda_2(\zeta_1,\eta_1,\zeta_2,\eta_2)= \lambda_2(\zeta_2, \eta_2)\in\mathbb{R}~,
\end{align}
\end{subequations}
such that $g_{\mu\nu}$ becomes Hermitian. We will come to the importance of this in a little bit. But for now, we study the effect of the elements of a group $G$, as linear operators, on a complex manifold and the coordinate transformations related to $G$. Such operations define a set of homomorphisms from $G$ to the general linear group $\GL(n,\mathbb{C})$, and such homomorphisms to the general linear group define an $n-$dimensional matrix representation. The matrix representation is useful when it works on any manifold chart, i.e. without fixing the manifold's basis. In that sense, a matrix representation of $G$ is a realization of $G$ elements as matrices affecting an $n-$dimensional complex space of column vectors. Additionally, the change of the manifold's basis results in conjugation of the matrix representation of $G$. Furthermore, a matrix representation on a manifold and a group operation on a manifold are two equivalent concepts. The later defines the group orbits and group stabilizers. It is interesting to study groups of Lie isometries and their symmetries of manifolds which the $G$ elements act transitively on. We define the isotropy group as $G_p = \{g \in G,\ gp = p\},\ p \in \mathcal{M}$, and the orbit of $G$ through $p$ by $Gp = \{gp,\ g \in G\} \simeq G/G_p$. And an orbit becomes a stabilizer if $G\equiv G_p$ at $p\in\mathcal{M}$.\\
\noindent
The transformations~(\ref{u2=r-2m})--(\ref{sheets C-transformation}) show that a matrix $g_{\mu\nu}$ should belong to the general linear group $\GL(4,\mathbb{R}):=\{T\in\M_4(\mathbb{R}):\det T\neq 0\}$, where $\M_4(\mathbb{R})$ is the space of all real $4\times 4$ matrices.
We can exploit the bijective relation $\GL(n,\mathbb{C})\leftrightarrow\GL(2n,\mathbb{R})$ to complexify the spacetime. Without loss of generality, we try $n=2$ such that the last relation means $Z\mapsto T:=\mathbb{R}Z$ on the elementary level for the complex matrix $Z\in \M_2(\mathbb{C})$. This means the $2\times 2$ complex matrices $Z$ can be characterized as $4\times 4$ real matrices $T$ such that they preserve the action of the linear complex structure $J:\mathcal{M}\to \mathcal{M}$ on the metric and the manifold. The complex structure is characterized by $J^2=-I$ for a manifold $\mathcal{M}$ upon which $Z$ acts. It is worth noting that the action of $J$ on $\mathcal{M}$ complexifies the tangent bundle $T\mathcal{M}^{{\mathbb{C}}}$ and introduces the conjugate tangent bundle too.\\
\noindent
So to complexify spacetime, we need to construct the conjugate group $T^{\dagger}HT:=\SU(2)\cap \mathcal{M}$, where $H$ is a $2\times 2$ Hermitian complex form of $g_{\mu\overline{\nu}}=h_{\mu\nu}+ik_{\mu\nu}$, i.e. $g_{\mu\nu}=g_{\overline{\mu}\overline{\nu}}$, and SU(2) is the special unitary subgroup. This guarantees the invariance of the Hermitian form
\begin{equation}
    \langle T\zeta, T\eta\rangle=\langle \zeta, \eta\rangle=\zeta_1\zeta_2-\eta_1\eta_2~.
\end{equation}
We know that some $Z\in\GL(2,\mathbb{C})$ can be defined as the \emph{special linear subgroup} $\SL(2,\mathbb{C}):=\{T\in\GL(2,\mathbb{C}): \det T=1\}$. Moreover, there exists another subgroup known as the \emph{unitary subgroup} $\U(2):=\{T\in\GL(2,\mathbb{C}): T^{\dagger}I_{(1,1)}T=I_{(1,1)}\}$, where $T^{\dagger}$ is the conjugate transpose of $T$ and $I_{(1,1)}=diag(1,-1)$ as in the previously mentioned Hermitian form. Finally, the compact Lie \emph{special unitary subgroup} is defined as $\SU(2):=\U(2)\cap\SL(2,\mathbb{C})$. The importance of unitary subgroups stems from the textbook fact that every finite subgroup of $\GL(2,\mathbb{C})$ is \emph{conjugate} to a subgroup of $\U(2)$, and the proof is based on the $\GL$-invariance, that is, the unitary representation preserves the length of any vector belonging to $\mathcal{M}$. If we restrict $H$ subgroup to be the diagonal matrices in $\SU(2)\subset\M_2(\mathbb{C})$, the cosets $T^{\dagger}H$ would \emph{partition} the manifold associated with $\SU(2)$. We will see the importance of this when we reach the process of \emph{Hopf fibration}.\\
\noindent
Now, a Lie topological group $G$, including $\SU(2)$, acts \emph{continuously} on $\mathcal{M}$ by a set of homeomorphisms $\Phi:G\times\mathcal{M}\rightarrow\mathcal{M}; (g,p)\to\Phi_p(g)=gp$. This action is called \emph{proper} for any compact group. That is, any proper map inside $\Psi:G\times\mathcal{M}\rightarrow\mathcal{M\times\mathcal{M}}$, such that $(g,p)\to\Psi(g)(p)=(gp,p)$, should have a compact inverse. Since the isotropy group $\SU(2)_p$ is compact, then its representation $\chi_p:\SU(2)_p\rightarrow\GL(2,T_p\mathcal{M}^{\mathbb{C}})$ is continuous, where $\chi_p\in \text{Is}_p(g)$ the linear isotropy of the group element $g$. Such representation sends the group elements $g$ into their diffeomorphic actions $d_g\in\text{Diff}(\mathcal{M})$ on $\mathcal{M}$, where $d_g:=\sqrt{ds^2}$ is the linear isotropy of the group element $\text{Is}_p(g)$ associated with the invariant distance or the manifold metric \cite{helgason1979differential}.\\
\noindent
A manifold $\mathcal{M}$ is biholomorphically equivalent to $\mathbb{C}$ when the holomorphic automorphisms $\text{Aut}(\mathcal{M})$ of the manifold are isomorphic to $\text{Aut}(\mathbb{C})$. Then, the action of $\SU(2)$ identifies the rotationally
symmetric complex manifolds \cite{isaev2002effective}. We are interested in the case when the $\SU(2)$-orbit of $p$ is the orthogonal group $O_p$. In this case, and with the help of the conjugation of vectors by the complex structure, $T_p\mathcal{M}^{\mathbb{C}}$ can be split into $V\oplus iV$ for any $V\in T_p\mathcal{M}^{\mathbb{C}}$ \cite{flaherty1976Hermitian}. Therefore, $O_p$ becomes real hypersurface orbits of $\mathcal{M}$.\\
\noindent
For the sake of convenience, it is suggested to represent $\SU(2)$ action in terms of coordinate charts at every point like Eq.~(\ref{sheets C-transformation}). Now, the function $\varphi\colon\mathbb{C}^2\to \M_2(\mathbb{C})$ defined by\\
\noindent
\begin{equation}
\varphi(z_1,z_2)={
\begin{pmatrix}z_1 &-{\overline{z_2}}\\z_2 &{\overline{z_1}}\end{pmatrix}}~.
\end{equation}
verifies $\varphi(S^{3})=\SU(2)$, see for instance \cite[Example~16.9]{hall2013quantum}; and the details of finding the equivariant maps, that relate $q\in S^3$ to $p\in O_p$ of $\mathcal{M}$, and CR-diffeomorphism structure of $S^{3}$ are in Ref~\cite{isaev2002effective}. The equivariant diffeomorphism $f: S^3\rightarrow O_p$, where $f(g(p))=g(f(p))$, establishes the correspondence between the parameters $(\zeta_1,\eta_1,\zeta_2,\eta_2)$ of the point $p\in\mathcal{M}$ and the unitary action of $\SU(2)$ on $q=(z_i,z_2)\in S^3$ endowed from the fact that every unitary representation on a Hermitian vector space $V$ is a direct sum of the \emph{irreducible representations} of the group. This is crucial for finding the \emph{group orbit}, or \emph{congruence classes}, containing the conjugate subgroups of $\SU(2)$. It is known that setting one of the $z_i=0$ will define the other $z_j$ to be the longitude of $\SU(2)$ corresponding to the conjugate class $T^{\dagger}HT$, where $H\subset\SU(2)$. The equivariant diffeomorphism relates any $H$ as a metric $g_{\mu\nu}$ to a \emph{diagonalizeble} matrix $\mathcal{H}$ on $S^3$ using $\varphi(z_1,z_2)$, and a \emph{diagonalized} $\mathcal{H}$ can be read as
\begin{equation}
\mathcal{H}'=T^{\dagger}\mathcal{H}T={
\begin{pmatrix}\xi &0\\
0 &{\overline{\xi}}
\end{pmatrix}}~,
\end{equation}
where for any wormhole sheet we define $\xi=\kappa+i\lambda$ and $z_i=\alpha_i+i\beta_i,\ i=1,2$ such that
\be
T\mathcal{H}'T^{\dagger}=
\begin{pmatrix}
z_1 &-\overline{z_2}\\
z_2 &\overline{z_1}
\end{pmatrix}
\begin{pmatrix}
\xi &0\\
0 &\overline{\xi}
\end{pmatrix}
\begin{pmatrix}
\overline{z_1} &\overline{z_2}\\
-z_2 &z_1
\end{pmatrix}
=
\begin{pmatrix}
z_1\overline{z_1}\xi+z_2\overline{z_2}\overline{\xi} &~z_1\overline{z_2}\xi-z_1\overline{z_2}\overline{\xi}\\
~&~\\
\overline{z_1}z_2\xi-\overline{z_1}z_2\overline{\xi} &~z_1\overline{z_1}\overline{\xi}+z_2\overline{z_2}\xi
\end{pmatrix}~.
\ee
So, if we want to return back to the $\mathbb{R}^4$ space, and for any $\xi$, the last transformation ushers us to define the real vector $x:=(x^1,x^2,x^3,x^4)$ as
\begin{eqnarray}
x^1 &=& \kappa~,\\
x^2 &=& (\alpha^2_1+\beta^2_1-\alpha^2_2-\beta^2_2)\lambda~,\\
x^3 &=& 2(\alpha_1\beta_2+\alpha_2\beta_1)\lambda~,\\
x^4 &=& 2(\alpha_1\alpha_2-\beta_1\beta_2)\lambda~.
\end{eqnarray}
This means the complex geodesics on the sheet~1 and sheet~2 are endowed with a $\SU(2)$ symmetry, which is guaranteed by the conjugacy $T^{\dagger}HT$ that defines all longitudes of $\SU(2)$. Also, this may introduce a connection between external geometry and internal symmetry, it may show us a geometric origin/meaning for the unitary symmetry in physics\footnote{For a $\GL(2n,\mathbb{R})$ of $V$ or a Lorentz transformation on the flat Minkowski spacetime in particular, the group can be determined uniquely by its action on the null vectors that correspond to $S^2$.}. As we can notice, $\SU(2)$ symmetry for geodesics is local symmetry because the parameters $(\alpha_1,\beta_1,\alpha_2,\beta_2)$ depend on the position on wormhole. So, if the coordinates $\zeta$ and $\eta$ render geodesics
\begin{eqnarray}
\lambda_1&=&\vartheta_1(\alpha_2,\beta_2)~,\\
\lambda_2&=&\vartheta_2(\alpha_2,\beta_2)~,
\end{eqnarray}
where $\vartheta_1$ and $\vartheta_2$ are continuous functions in $x^i$, and $\alpha_1^2+\beta_1^2=1$, then $\SU(2)$ symmetry reduces to $\U(1)$. In that sense, $\SU(2)$ introduces local symmetry of complex vectors on the two sheets of wormholes, and $\U(1)$ symmetry introduces a local symmetry of the complex vector on the same sheet.\\
\noindent
It is worth noting that $S^3$ is diffeomorphic with $\SU(2)$ \cite[p.~127]{warner1983foundations}. Moreover, $S^3$ can be seen as a fiber bundle following the diagram
\be\label{Hopf}
\begin{matrix}
S^1\, \longrightarrow \, S^3\, \xrightarrow{\,\ \pi \ }\, S^2
\end{matrix}~,
\ee
with the Hopf fibration plotted in Figure~\ref{fig:Hopf} and used in physics in \cite{chang2011hopf,urbantke2003hopf} and for wormholes in \cite{dzhunushaliev2014kaluza}. 
\noindent
Now, we are ready to complexify the wormhole metric. First, we rearrange Eq.~(\ref{metric_S1}) such that it becomes
\be\label{rearr_wormhole_S}
ds^2=+\left[\left(1-\frac{2M}{r(\zeta)}\right)dt^2-(u^2+2M)^2\sin^2\theta d\phi^2\right]-\left[4(u^2+2M)du^2+(u^2+2M)^2d\theta^2\right]~.
\ee
Since we adopt the parameter $u(r)$ as defined in Eq.~(\ref{u2=r-2m}), we also define the complex parameter $\zeta$, its squared length, and any infinitesimal change in it as
\begin{eqnarray}
\zeta&=&\frac{1}{2}(u^2+2M)e^{i\theta}~,\\
\zeta\overline{\zeta}&=&\frac{(u^2+2M)^2}{4}=\frac{r^2(u)}{4}\label{zeta-zeta-bar}~,\\
d\zeta&=&\frac{1}{u^2+2M}u\zeta du+i\zeta d\theta\label{dzeta}~.
\end{eqnarray}
Eq.~(\ref{dzeta}) gives
\be
d\zeta d\overline{\zeta}=u^2du^2+\frac{1}{4}(u^2+2M)^2d\theta^2~,
\ee 
or
\be\label{dzeta-dzetabar+2Mdu2}
4(d\zeta d\overline{\zeta}+2Mdu^2)=4(u^2+2M)du^2+(u^2+2M)^2d\theta^2~.
\ee
Meanwhile Eq.~(\ref{zeta-zeta-bar}) yields
\be
(\zeta d\overline{\zeta}+\overline{\zeta} d\zeta)^2=u^2(u^2+2M)^2du^2=\left(2\sqrt{\zeta \overline{\zeta}}-2M\right)4(\zeta \overline{\zeta})du^2~,
\ee 
or
\be
du^2=\frac{(\zeta d\overline{\zeta}+\overline{\zeta} d\zeta)^2}{4(\zeta \overline{\zeta})\left(2\sqrt{\zeta \overline{\zeta}}-2M\right)}~.
\ee
Substitute the last result in the LHS of Eq.~(\ref{dzeta-dzetabar+2Mdu2}) to get
\be\label{4rdu2+r2dtheta2}
4\left[d\zeta d\overline{\zeta}+M\frac{(\zeta d\overline{\zeta}+\overline{\zeta} d\zeta)^2}{2(\zeta \overline{\zeta})\left(2\sqrt{\zeta \overline{\zeta}}-2M\right)}\right]=
4(u^2+2M)du^2+(u^2+2M)^2d\theta^2~.
\ee
In addition, the stereographic projection of $r(x,y,z)$ on the complex plane of $\zeta(\kappa,\lambda)$ with $\frac{x}{\kappa}=\frac{y}{\lambda}=1-z$ give \cite{ablowitz2003complex}
\be\label{stereographic}
\sin^2\theta=\frac{4\zeta\overline{\zeta}}{\left(1+\zeta\overline{\zeta}\right)^2}~.
\ee
Furthermore, set
\be
\eta=t+iM\phi
\ee
such that
\begin{eqnarray}
dt^2&=&\frac{1}{4}(d\eta+d\overline{\eta})^2\label{dt2}~,\\
d\phi^2&=&\frac{1}{4M^2}(d\eta-d\overline{\eta})^2\label{dphi2}~.
\end{eqnarray}
Finally, we substitute Eq.~(\ref{4rdu2+r2dtheta2},\ref{stereographic},\ref{dt2},\ref{dphi2}) to get
\be\label{metric_S2}
ds^2=\left(1-\frac{M}{\sqrt{\zeta\overline{\zeta}}}\right)(d\eta+d\overline{\eta})^2
-4\frac{(\zeta\overline{\zeta})^2}{M^2\left(1+\zeta\overline{\zeta}\right)^2}(d\eta-d\overline{\eta})^2
-4\left[d\zeta d\overline{\zeta}+M\frac{(\zeta d\overline{\zeta}+\overline{\zeta} d\zeta)^2}{2(\zeta \overline{\zeta})(2\sqrt{\zeta \overline{\zeta}}-2M)}\right]
\ee
which is not yet a \textit{manifestly} Hermitian metric despite being a general $2-$dimensional metric of such complex manifold.\\
In order to make the metric (\ref{rearr_wormhole_S}) Hermitian, we need to consider the following coordinate redefinition 
\begin{equation}
d\tilde{u}=\frac{2(u^2+2M)}{u}du~,
\end{equation}
or
\be
\tilde{u}=\int^{u}_{-2M}\frac{2(u^2+2M)}{u}du
=u^2-4M^2+4M\ln\left(-\frac{u}{2M}\right)=u^2-4M^2+4M\ln\left(\frac{u}{2M}\right)+i\pi
\ee
where $\displaystyle u^2-4M^2+4M\ln\left(\frac{u}{2M}\right)\in\mathbb{R}$. Then, the metric (\ref{rearr_wormhole_S}) becomes
\begin{equation}
    ds^2=-\left[\frac{u^2(\tilde{u})}{u^2(\tilde{u})+2M}(dt^2-d\tilde{u}^2)+(u^2(\tilde{u})+2M)^2d\Omega^2\right]~,
\end{equation}
which is not Hermitian yet as $\tilde{u}\in\mathbb{C}$. However, $e^{\tilde{u}}\in\mathbb{R}$ indeed. In order to improve the previous metric into a Hermitian, we use the following Rindler-like coordinates together with Wick rotation and the complex coordinates
\begin{subequations}
\begin{align}
    X&=e^{\tilde{u}}\cosh t~,\\
    T&=e^{\tilde{u}}\sinh t,\ T\rightarrow iT~,\\
    \eta&=X+iT~,\\
    d\theta^2+\sin^2\theta d\phi^2&=\frac{d\zeta d\overline{\zeta}}{\left(1+\frac{1}{4}\zeta \overline{\zeta}\right)^2}~,
\end{align}
\end{subequations}
such that the relevant metric becomes
\begin{equation}\label{HermitianRindler}
    ds^2=h(\eta+\overline{\eta})d\eta d\overline{\eta}+k(\eta+\overline{\eta})\frac{d\zeta d\overline{\zeta}}{\left(1+\frac{1}{4}\zeta \overline{\zeta}\right)^2}~,
\end{equation}
which is an Euclidean metric of the corresponding $2-$dimensional Hermitian complex manifold for arbitrary real valued functions $h(\eta+\overline{\eta})$ and $k(\eta+\overline{\eta})$, see \cite[pages~44-45]{rajan2015complex} for more details. There are many other ways to render a Hermitian metric. Whether the metric is real or Hermitian, the process of complexification visualizes how Schwarzschild wormholes behave in the realm of complex geometry. This is an important result as it could help studying the \emph{wavefunction of wormholes} upon analyzing the geometry as a Quantum Field Theory (QFT) in complex curved spacetime \cite{Barvinsky1995the}.

\begin{remark}
Consider the de Sitter-Schwarzschild metric
\begin{equation}
ds^2= -\left({1-\frac{2M}{r}}-\frac{\Lambda}{3}r^2\right)^{-1} dr^2- r^2\left(d\theta^2+\sin^2\theta d\phi^2\right) +\left(1-\frac{2M}{r}-\frac{\Lambda}{3}r^2\right)dt^2~,
\end{equation}
where $\Lambda>0$. By using the results about depressed cubic equations given for instance in \cite{neumark1965solution}, the polynomial $f(r)=-\frac{\Lambda}{3}r^3+r-2M$ has three real roots if and only if $\Lambda<\frac{1}{9M^2}$. If $\Lambda>\frac{1}{9M^2}$, then $f(r)$ has one real root and two complex conjugate roots. In wormhole geometry, real horizon means the possibility to measure beyond it, and complex horizon means the impossibility to measure beyond it.
\end{remark}

\section{Reissner-Nordström wormhole geometry}\label{Sec:RN}
\noindent
In order to understand the geometric origin of the charge, Einstein and Rosen \cite{einstein1935particle} investigated the following exotic Reissner-Nordström metric
\be
ds^2=-\left(1-\frac{2M}{r}-{\frac {Q^2}{r^2}}\right)^{-1}dr^2-r^2\left(d\theta^2+\sin^2\theta\,d\phi^2\right)+\left(1-\frac{2M}{r}-\frac{Q^2}{r^2}\right)dt^2~,
\ee
where $M>0$ and $Q>0$ for exotic matter with negative energy density. If we consider the following transformation 
\be
u^2=r^2-2 Mr-Q^2~,
\ee
it leads to $u^2du^2=(r-M)^2 dr^2$. In the new $u$ coordinate system, one obtains for $ds^2$ the expression
\be
ds^2=-\frac{r^2}{(r-M)^2}du^2
-\left(u^2+2Mr+Q^2\right)\left(d\theta^2+\sin^2\theta\,d\phi^2\right)+\frac{u^2}{u^2+2Mr+Q^2}dt^2~.
\ee
We have $(r-M)^2=u^2+M^2+Q^2$. Consider the continuous and positive function $u\mapsto \tilde{f}(u)$ defined by
\be
r=\sqrt{u^2+M^2+Q^2}+M:=\tilde{f}(u)~,
\ee
and one obtains for $ds^2$ the expression
\be\label{metric_ER1}
ds^2=-\frac{\tilde{f}(u)^2}{u^2+M^2+Q^2}du^2
-\left(u^2+2M\tilde{f}(u)+Q^2\right)\left(d\theta^2+\sin^2\theta\,d\phi^2\right)+\frac{u^2}{u^2+2M\tilde{f}(u)+Q^2}dt^2
\ee
We find the coordinate $u$ vanishes at the event horizons when $r_1=M-\sqrt{M^2+Q^2}$ and $r_2=M+\sqrt{M^2+Q^2}$. In the $u$ coordinate, the bridge at $r=r_2$ verifies $r_1<0<M<r_2$. The metric~\eqref{metric_ER1} is defined properly until $r=M$ and the singularity at $r=r_2$ is removed. So, we obtain two regions for the first sheet:
\begin{itemize}
\item $u$ has imaginary value when $r$ varies from $0$ to $r_2$;
\item $u$ has real value from $0$ to $+\infty$ when when $r$ varies from $r_2$ to $+\infty$.
\end{itemize}
Similarly, we have two regions for the other sheet. When $0<r<M$, the function $\tilde{f}(u)$ in the metric~\eqref{metric_ER1} must be replaced by $-\sqrt{u^2+M^2+Q^2}+M$. As in \cite{einstein1935particle} and for sake of simplicity, we consider that $M=0$. In that case, the metric~\eqref{metric_ER1} reduces to 
\begin{equation}\label{metric_ER2}
ds^2=-du^2-(u^2+Q^2)(d\theta^2+\sin^2\theta d\phi^2)+\frac{u^2}{u^2+Q^2}dt^2.
\end{equation}
It is possible to obtain a metric very similar to \eqref{HermitianRindler} by using similar calculations, except that $h$ and $k$ become functions in $Q$ but not in $M$. Calculations are left to the reader.

\begin{remark}
Fist, consider the classical Reissner-Nordström metric
\be\label{metric_RN}
ds^2=-\left(1-\frac{2M}{r}+{\frac {Q^2}{r^2}}\right)^{-1}dr^2-r^2\left(d\theta^2+\sin^2\theta\,d\phi^2\right)+\left(1-\frac{2M}{r}+\frac{Q^2}{r^2}\right)dt^2,
\ee
where $M>0$ and $Q>0$. We choose that 
\be\label{Condition+}
M>Q>0~. 
\ee
Consider the following transformation
\be
u^2=r^2-2 Mr+Q^2~,
\ee
which gives $u^2du^2=(r-M)^2 dr^2$. In the new $``u"$ coordinate system, one obtains for $ds^2$ the expression
\be
ds^2=-\frac{r^2}{(r-M)^2}du^2-\left(u^2+2Mr-Q^2\right)\left(d\theta^2+\sin^2\theta\,d\phi^2\right)+\frac{u^2}{u^2+2Mr-Q^2}dt^2~.
\ee
We have $(r-M)^2=u^2+M^2-Q^2$. For $r>M$, and by using condition~\eqref{Condition+}, we obtain
\be
r=\sqrt{u^2+M^2-Q^2}+M:=\tilde{g}(u)~,
\ee
with $u\mapsto \tilde{g}(u)$ continuous and positive. In the new coordinate system, one obtains for $ds^2$ the expression
\be\label{wormhole_RN}
ds^2=-\frac{\tilde{g}(u)^2}{u^2+M^2-Q^2}du^2-\left(u^2+2M\tilde{g}(u)-Q^2\right)\left(d\theta^2+\sin^2\theta\,d\phi^2\right)+\frac{u^2}{u^2+2M\tilde{g}(u)-Q^2}dt^2
\ee
We find the coordinate $u$ vanishes at the event horizons when $r_1=M-\sqrt{M^2- Q^2}$ and $r_2=M+\sqrt{M^2-Q^2}$. In the $u$ coordinate, the bridge at $r=r_2$ verifies $0<r_1<M<r_2$. The metric~\eqref{wormhole_RN} is defined until $r=M$ and the singularity at $r=r_2$ is removed. So, we obtain three regions for the first sheet:
\begin{itemize}
\item $u$ has real value from $Q^2$ to $0$ when $r$ varies from $0$ to $r_1$;
\item $u$ has imaginary value when $r$ varies from $r_1$ to $r_2$;
\item $u$ has real value from $0$ to $+\infty$ when when $r$ varies from $r_2$ to $+\infty$.
\end{itemize}
We also have three regions for the other sheet. When $0<r<M$, the function $\tilde{g}(u)$ must be replaced by $-\sqrt{u^2+M^2-Q^2}+M$ in the metric~\eqref{wormhole_RN}. The situation is therefore different from those presented previously and additional studies will be necessary.\\
Then, we know that $\SU(3)$ follows the diagram
\be\label{SU(3)}
\begin{matrix}
SU(2)\, \longrightarrow \, SU(3)\, \xrightarrow{\,\ \pi \ }\, S^5~,
\end{matrix}
\ee
where $\pi$ is the projection, see for instance \cite[Proposition~13.11]{hall2015lie}. The special unitary group $\SU(3)$ is the nontrivial $\SU(2)-$bundle over $S^5$, see for instance \cite[Section~3]{lafont2019sets}. Moreover, $S^5$ is diffeomorphic with $\SU(3)/\SU(2)$ \cite[p.~127]{warner1983foundations}. However, the way of building an Euclidean metric on a complex Hermitian
manifold involving the $\SU(3)$ symmetry is an open problem. We also notice that if $M^2\leq Q^2$, then the polynomial $P(r)=r^2-2M+Q^2$ is always positive such that the change of variable $u$ cannot provide unitary symmetries. Contrary to the Schwarzschild and exotic Reissner-Nordström wormhole geometry, the classical Reissner-Nordström wormhole geometry implies the mass-charge condition~\eqref{Condition+} which is also used to avoid naked singularities \cite[Section~12.6]{hobson2006general}.
\end{remark}

\section{Quantum tunneling and wormhole thermodynamics}\label{Sec:tunneling}
\noindent
Discovering unitary symmetries in wormhole geometry motivates us to explore the quantum properties of wormholes. Being traversable for a wormhole is a challenge, see for instance \cite{morris1988wormholes,morris1988wormholes2}. Wormholes are generally non-traversable for classical matter \cite{kruskal1960maximal} but they can be modified to be traversable by removing event horizons, see \cite{hayward2004make,poisson1995thin,simpson2019black,visser1989traversable} for Schwarzschild-like wormholes and \cite{blazquez2021traversable,lobo2016flamm} for Reissner-Nordström-like wormholes. We know that particles are subject to quantum tunneling, which makes Schwarzschild and Reissner-Nordström wormholes traversable for particles while keeping event horizons. A similar idea has been used since the seminal works of Bekenstein in \cite{bekenstein1972black} and Hawking in \cite{hawking1975particle} for studying the black hole radiation. In this section, we develop quantum tunneling and wormhole thermodynamics by computing the Hawking temperature.

\subsection{Schwarzschild wormhole case}
\noindent
First, we point out and observe an interesting fact about the radial null curves in the wormhole metric~\eqref{metric_S1} by setting $ds^2=d\theta=d\phi=0$, yielding
\be
\frac{du}{dt}=\pm \frac{u}{2 (u^2+2M)}~.
\ee
The above quantity defines the “coordinate speed of light” for the wormhole metric, and as we can see there is a horizon with a coordinate location $u=0$ yielding 
\be
\left.\frac{du}{dt}\right|_{u=0}\longrightarrow 0~.
\ee
The presence of the horizon implies that the quantum tunneling of particles from ``another universe'' to our universe can form Hawking radiation and, consequently, detecting particles by a distant observer located in our universe. We can study the tunneling of different massless or massive spin particles; and in the present work, we focus on studying the tunneling of vector particles. The motion of a massive vector particle of mass $m$, described by the vector field
$\psi^{\mu}$, \emph{might} be studied by the Proca equation (PE), which reads \cite{Illge:1993up}
\be\label{Proca}
\nabla_{\mu}\nabla^{[\mu}\psi^{\nu]}-\frac{m^{2}}{\hbar^{2}}\psi^{\nu}=\frac{1}{\sqrt{-g}}\partial_{\mu}\left[\sqrt{-g}\partial^{[\mu}\psi^{\nu]}\right]-\frac{m^{2}}{\hbar^{2}}\psi^{\nu}=0
\ee
where, from the metric~\eqref{metric_S1}, we define the determinant $\sqrt{-g}=2u^2(u^2+2M)^2 \sin\theta$, and 
\be
\nabla_{[\mu}\psi_{\nu]}=\frac{1}{2}(\nabla_{\mu}\psi_{\nu}-\nabla_{\nu}\psi_{\mu}):=\psi_{\mu\nu}~.
\ee
The corresponding action is
\be
    S=-\int d^4x\sqrt{g}\left( \frac{1}{2}\psi_{\mu\nu}\psi^{\mu\nu}+\frac{m^2}{\hbar^2}\psi_{\mu}\psi^{\mu}\right)~.
\ee
Then in any curvilinear coordinates, and using the Bianchi-Ricci identity $\nabla_{[\lambda}\psi_{\mu\nu]}=0$, we get the true version of Eq.~(\ref{Proca}) as a QFT in curved spacetime equations of motion
\begin{subequations}
\begin{align}
    &\nabla^{\nu}\nabla_{[\nu}\psi_{\mu]}-\mathcal{R}_{\mu}^{~\nu}\psi_{\nu}-\frac{m^{2}}{\hbar^{2}}\psi_{\mu}=0~,\\
    &\nabla_{\lambda}\nabla^{\lambda}\psi_{\mu\nu}+\mathcal{C}_{\mu\nu}^{~~\kappa\lambda}\psi_{\kappa\lambda}-\left(\frac{\mathcal{R}}{3}+\frac{m^2}{\hbar^2}\right)\psi_{\mu\nu}=0~,
\end{align}
\end{subequations}
where $\mathcal{R}_{\mu\nu}$ and $\mathcal{R}$ are the Ricci tensor and the Ricci scalar respectively, and $\mathcal{C}_{\mu\nu\kappa\lambda}$ is the trace-free conformal curvature tensor. Taking the flat limit $g_{\mu\nu}\to\eta_{\mu\nu}$ changes the essence of the last two equations to become Lorentz invariant.\\
\noindent
Solving tunneling equations exactly is quite hard. So, we apply the WKB approximation method 
\be\label{WKB}
\psi_{\nu}=C_{\nu}(t,u,\theta,\phi)e^{\left(\frac{i}{\hbar}\left(S_{0}(t,u,\theta,\phi)+\hbar\,S_{1}(t,u,\theta,\phi)+\hdots\right)\right)}~.
\ee
Taking into the consideration the symmetries of the metric~\eqref{metric_S1} given
by three corresponding Killing vectors $(\partial/\partial_{t})^{\mu}$ and
$(\partial/\partial_{\phi})^{\mu}$,
we may choose the following ansatz for the action 
\be\label{Action}
S_{0}(t,u,\theta,\phi,\psi)=Et+R(r,\theta)-j\phi~,
\ee
where $E$ is the energy of the particle, and $j$ and $l$ denotes the angular momentum of the particle corresponding to the angles $\phi$ and $\psi$, respectively. If we keep only the leading order of $\hbar$, we find a set of four differential equations. These equations can help us to construct a $4\times4$ matrix $\aleph$, which satisfies the following matrix equation 
\be
 \aleph(C_{1},C_{2},C_{3},C_{4})^{T}=0~.
\ee
We solve for the radial part to get the following integral 
\be
R_{\pm}=\pm \int \frac{2 \sqrt{E^2-\frac{u^2}{u^2+2M}\left[m^2+\Delta(u) \right]}} {\mathcal{F}(u)}du~,
\ee
where 
\be
\Delta(u)=\frac{(\partial_{\theta}R)^2}{(u^2+2M)^2}+\frac{j^2}{(u^2+2M)^2 \sin^2\theta}~,
\ee
and 
\be
\mathcal{F}(u)=\frac{u}{u^2+2M}=\mathcal{F}'(u)|_{u=0} (u-u_h)+\cdots
\ee
Now, there is a singularity in the above integral when $u_h=0$, meaning that $\mathcal{F}\to 0$. So in order to find the Hawking temperature, we now make use of the equation
\be
\lim_{\epsilon \to 0} \text{Im}\frac{1}{u-u_h\pm i \epsilon }=\delta(u-u_h)~,
\ee
where $u_h=0$. In this way we find 
\be
\text{Im}R_{\pm}=\pm \frac{2 E \pi }{\mathcal{F}'(u)|_{u=0}}~.
\ee
Using $p_u^{\pm}=\pm \partial_u R_{\pm}$, for the total tunneling rate gives
\be
\Gamma=\exp\left(-\frac{1}{\hbar}\text{Im} \oint p_{u} \mathrm{d}r\right)=\exp\left[-\frac{1}{\hbar} \text{Im} \left(\int p_{u}^{+}\mathrm{d}u-\int p_{u}^{-}\mathrm{d}u\right) \right]= \exp\left(-\frac{4 E \pi }{\hbar \mathcal{F}'(u)|_{u=0}}\right).
\ee
It is interesting that, for the black hole case, there is a temporal part contribution due to the connection of the interior region and the exterior region of the black hole. In the wormhole case, we don't have such a contribution. We can finally obtain the Hawking temperature for the wormhole by using the 
Boltzmann factor $\Gamma=\exp(-E/T)$, and setting $\hbar$ to unity, so that it results with 
\be\label{Hawking_temperature_S}
    T=\frac{\mathcal{F}'(u)|_{u=0}}{4 \pi }=\frac{1}{8 \pi M}~.
\ee
This is interesting result as it shows that the Hawking temperature for the Schwarzschild wormhole coincides with the Schwarzschild black hole temperature. We can verify the above result for the Hawking temperature using a topological method based on the Gauss-Bonnet theorem reported in Ref~\cite{Robson:2018con,Robson:2019yzx}. Let's now rewrite the metric~\eqref{metric_S1} in a form of $2-$dimensional Euclidean spacetime given by
\be\label{wormhole_S2}
    ds^2=4(u^2+2M)du^2+\frac{u^2}{u^2+2M}d\tau^2~.
\ee
The Hawking temperature can be found from \cite{Robson:2018con}
\be\label{Integral}
T=\frac{\hbar c}{4 \pi \chi k_B}\sum_{j \leq \chi}\int_{u_h} \sqrt{g}\, \mathcal{R}\, du~.
\ee
Applying this equation for the wormhole metric~\eqref{wormhole_S2}, we find the Ricci scalar 
\be
\mathcal{R}=\frac{4M}{(u^2+2M)^3}
\ee
and $\sqrt{g}=2 u$. Setting $\hbar=c=k_B=1$, using the fact that the Euler characteristic of Euclidean geometry is $\chi=1$ at the wormhole horizon $u_h=0$, we solve the integral~\eqref{Integral} and obtain 
\be
T=\frac{1}{4 \pi}\int_{0}^{\infty} \frac{4M}{(u^2+2M)^3} 2\, u\, du=\frac{1}{8 \pi M}~.
\ee
which coincides with the Hawking temperature~\eqref{Hawking_temperature_S} obtained via tunneling. 

\subsection{Reissner-Nordström wormhole case}
\noindent
Here we shall consider a tunneling from massless RN wormhole geometry using metric~\eqref{metric_ER2}. For the radial null curve, and by setting $ds^2=d\theta=d\phi=0$, we obtain
\be
    \frac{du}{dt}=\pm \frac{u}{\sqrt{u^2+Q^2}}~,
\ee
and therefore we see that the points $u=0$ play the role of the horizon as $du/dt\to0$, provided that $u=0$. This indicates that there could be a quantum tunneling associated with the horizon. To find the Hawking temperature, we can apply the WKB approximation given by Eq.~\eqref{WKB} along with the action \eqref{Action}. Consequently, we construct a $4\times 4$ matrix 
\be
 \mathrm{M}(D_{1},D_{2},D_{3},D_{4})^{T}=0~,
\ee
where, for the radial part, we get the following integral 
\be
R_{\pm}=\pm \int \frac{\sqrt{E^2-\frac{u^2}{Q^2+u^2} \left[m^2+\xi(u) \right]}}{\mathcal{G}(u)} du~,
\ee
with 
\be
\xi(u)=\frac{(\partial_{\theta}R)^2}{(Q^2+u^2)^2}-\frac{j^2}{\sin^2\theta (Q^2+u^2)^2}~,
\ee
and 
\be
\mathcal{G}(u)=\frac{u}{\sqrt{Q^2+u^2}}=\mathcal{G}'(u)|_{u=0} (u-u_h)+\cdots
\ee
Now, there is a singularity in the above integral when $u_h=0$, meaning that $\mathcal{G}\to 0$. In order to find the Hawking temperature at $u_h=0$, we consider 
\be
\text{Im}R_{\pm}=\pm \frac{ E \pi }{\mathcal{G}'(u)|_{u=0}}~.
\ee
Using $p_u^{\pm}=\pm \partial_u R_{\pm}$, the total tunneling rate gives
\be
\Gamma =\exp\left(-\frac{2 E \pi }{\hbar \mathcal{G}'(u)|_{u=0}}\right)~.
\ee
Boltzmann factor $\Gamma=\exp(-E/T)$ leads to define the temperature as 
\be\label{Hawking_temperature_RN}
    T=\frac{\mathcal{G}'(u)|_{u=0}}{2 \pi }=\frac{1}{2 \pi Q}~.
\ee
Let's now derive the Hawking temperature using a topological method based on the Gauss-Bonnet theorem. To do so, we need to rewrite the metric~\eqref{metric_ER2} in a form of $2-$dimensional Euclidean spacetime given by
\be
    ds^2=du^2+\frac{u^2}{u^2+Q^2}d\tau^2~.
\ee
For the Ricci scalar, we obtain
\be
\mathcal{R}=\frac{6 Q^2}{(Q^2+u^2)^2}~,
\ee
and $\sqrt{g}=u (u^2+Q^2)^{-1/2}$. At the wormhole horizon $u_h=0$, we obtain 
\be
T=\frac{1}{4 \pi}\int_{0}^{\infty}\frac{6\, u\,Q^2}{(Q^2+u^2)^{5/2}} du=\frac{1}{2 \pi \,Q}~,
\ee
which coincides with the Hawking temperature~\eqref{Hawking_temperature_RN} obtained via tunneling.

\section{Concluding remarks}\label{Sec:conclusion}
\noindent
We closely looked at Schwarzschild and Reissner-Nordström wormhole geometry and obtained unitary symmetries $\U(1)$ and $\SU(2)$ using spacetime complexification; the study brings that attention to the possibility that wormholes could illustrate the relation between unitary symmetries and spacetime geometry. Additionally, we developed wormhole thermodynamics for Schwarzschild and Reissner-Nordström wormholes through quantum tunneling. The results are consistent with those of Hawking and Bekeinstein for black hole thermodynamics. It implies that particles can cross these wormholes. This could be related to the ER=EPR conjecture \cite{maldacena2013cool} and to the new experimental findings obtained from studying traversable wormholes/EPR pair entanglement within quantum computing regimes \cite{Jafferis2022}. We hope to report on these important results in the future.

\section*{Acknowledgement}
\noindent
AFA  would like to thank Shaaban Khalil and Ammar Kassim for the  discussions. K.J would like to thank Dejan Stojkovic for interesting comments during the preparation of this work. 

\bibliographystyle{apsrev4-1}
\bibliography{wormhole}
\end{document}